\documentclass{article}

  \usepackage{driftChamberML-defs}
  \usepackage[final]{neurips_2025_ml4ps}
  \usepackage{graphicx}
   \usepackage{amsmath} 

\usepackage[utf8]{inputenc} 
\usepackage[T1]{fontenc}    
\usepackage{hyperref}       
\usepackage{url}            
\usepackage{booktabs}       
\usepackage{amsfonts}       
\usepackage{nicefrac}       
\usepackage{microtype}      
\usepackage{xcolor}         

\title{Edge Machine Learning for Cluster Counting in Next-Generation Drift Chambers}

\author{%
  Deniz Yilmaz \\ 
  Department of Physics\\
  Stanford University\\
  450 Jane Stanford Way,  Stanford, CA 94305, United States of America \\
  \texttt{dyilmaz@stanford.edu } \\
  \And
  Liangyu Wu \\
  Department of Physics\\
  Stanford University\\
  450 Jane Stanford Way,  Stanford, CA 94305, United States of America \\
  \texttt{liangyu.wu@stanford.edu} \\
  \AND
  Julia Gonski \\
  SLAC National Accelerator Laboratory\\
  2575 Sand Hill Road, Menlo Park, CA 94025, United States of America\\
  \texttt{jgonski@slac.stanford.edu} \\
  \AND
  Dylan Rankin \\
  University of Pennsylvania \\
  209 S 33rd St, Philadelphia, PA 19104, United States of America\\
  \texttt{dsrankin@sas.upenn.edu} \\
    \AND
  Christian Herwig \\
  University of Michigan \\
  450 Church St, Ann Arbor, MI 48109, United States of America\\
  \texttt{herwig@umich.edu} \\
}

\begin{document}

\maketitle

\begin{abstract}
Drift chambers have long been central to collider tracking, but future machines like a Higgs factory motivate higher granularity and cluster counting for particle ID, posing new data processing challenges.
Machine learning (ML) at the "edge", or in cell-level readout, can dramatically reduce the off-detector data rate for high-granularity drift chambers by performing cluster counting at-source. 
We present machine learning algorithms for cluster counting in real-time readout of future drift chambers. 
These algorithms outperform traditional derivative-based techniques based on achievable pion-kaon separation. 
When synthesized to FPGA resources, they can achieve latencies consistent with real-time operation in a future Higgs factory scenario, thus advancing both R\&D for future collider detectors as well as hardware-based ML for edge applications in high energy physics. 
\end{abstract}

\section{Introduction}

Drift chambers are gaseous detectors that track charged particles with high precision and aid in particle identification (PID). As particles pass through the gas, primary ionization electrons drift under an electric field toward sense wires, where arrival times reveal the particle’s position; secondary electrons can form through further ionization. With fine spatial resolution, wide acceptance, and low material budget, drift chambers have long been central to collider experiments, including
Belle at KEK~\cite{ABASHIAN2002117} and BaBar at SLAC~\cite{BaBar:1995bns}, among others.

Recent strategic planning processes conducted by the high energy physics community have pointed to the importance of a future $e^+e^-$ Higgs factory in achieving prioritized future physics goals of the field~\cite{p5_2023}.
The low mass and high precision of drift chambers make them attractive for future $e^+e^-$ trackers, but adapting them to unprecedented luminosities poses novel challenges. 
High occupancies complicate signal separation, while long drift times limit performance at rapid bunch crossings. 
In addition, stringent demands on tracking and PID performance call for high-granularity designs that generate unprecedented off-detector data rates (towards TB/s)~\cite{Tassielli:2021rjk}. 

The introduction of machine learning (ML) methods to drift chamber data analysis can help address these challenges and enable performant and realizable next-generation drift chambers. 
Specifically, this work focuses on ML-assisted \textit{cluster counting} ($dN/dx$) for PID~\cite{Rolandi:2008ujz}. 
Traditionally, drift chambers have done PID using the average energy loss from ionization over length ($dE/dx)$ ~\cite{Allison:1980vw}, which suffers from large Landau fluctuations due to secondary electrons, thereby limiting the ultimate resolution for particle separation.
$dN/dx$ alternatively quantifies the number of primary ionization clusters per unit length, which follows a Poisson-like distribution, thus improving statistical resolution for more accurate PID measurements. 
Implementing $dN/dx$ in practice is difficult (especially at high track densities), as it requires fine-grained time and spatial resolution to distinguish closely space individual clusters, readout of single-ionization electron signal, and a gas mixture with favorable ionization and drift properties. 
Previous works have studied both the feasibility of $dN/dx$ at future $e^+e^-$ colliders~\cite{yu2025evaluationpidperformancecepc} and the ability of ML to improve $dN/dx$ using drift chamber simulation~\cite{ZHAO2024109208, Tian_2025}.

This work is the first to demonstrate ML methods that aim to simultaneously address the data rate and $dN/dx$ challenges of future drift chamber designs, by performing cluster counting at the ``edge". 
This feature extraction enables a dramatic reduction of the off-detector data rate by transmitting only the cluster count per waveform, while enhancing PID with respect to traditional methods. 
While a valuable proof-of-concept, this approach is limited by its use of a toy detector design, lack of precise data rate and power estimates needed for a full readout design, and full reliance on simulation, all of which would need to be addressed in subsequent works to fully realize and deploy edge ML in a future experiment.

\section{Methodology}



This study utilizes the dataset from Ref.~\cite{Tian_2025}, generated through Garfield++~\cite{garfield} simulations based on the fourth conceptual detector design for the Circular Electron Positron Collider (CEPC)~\cite{thecepcstudygroup2018cepcconceptualdesignreport}. 
The simulation employs a 1.5~GHz sampling rate with square drift cells of 18~mm $\times$ 18~mm. 
Each drift cell features a central sense wire surrounded by eight field wires in a square geometry. 
The detector operates with a gas mixture of 90\% He and 10\% iC$_4$H$_{10}$ to achieve optimal primary ionization density. The training dataset comprises $5 \times 10^5$ $\pi^{\pm}$ events with momenta ranging from 0.2 to 20.0~GeV/c 
, along with pion and kaon samples with $10^5$ events for each of seven momenta ranging from [5, 20] GeV at 2.5 GeV intervals for evaluating pion-kaon separation performance.

\begin{figure}[!htbp]
\centering
   \includegraphics[width=0.6\textwidth]{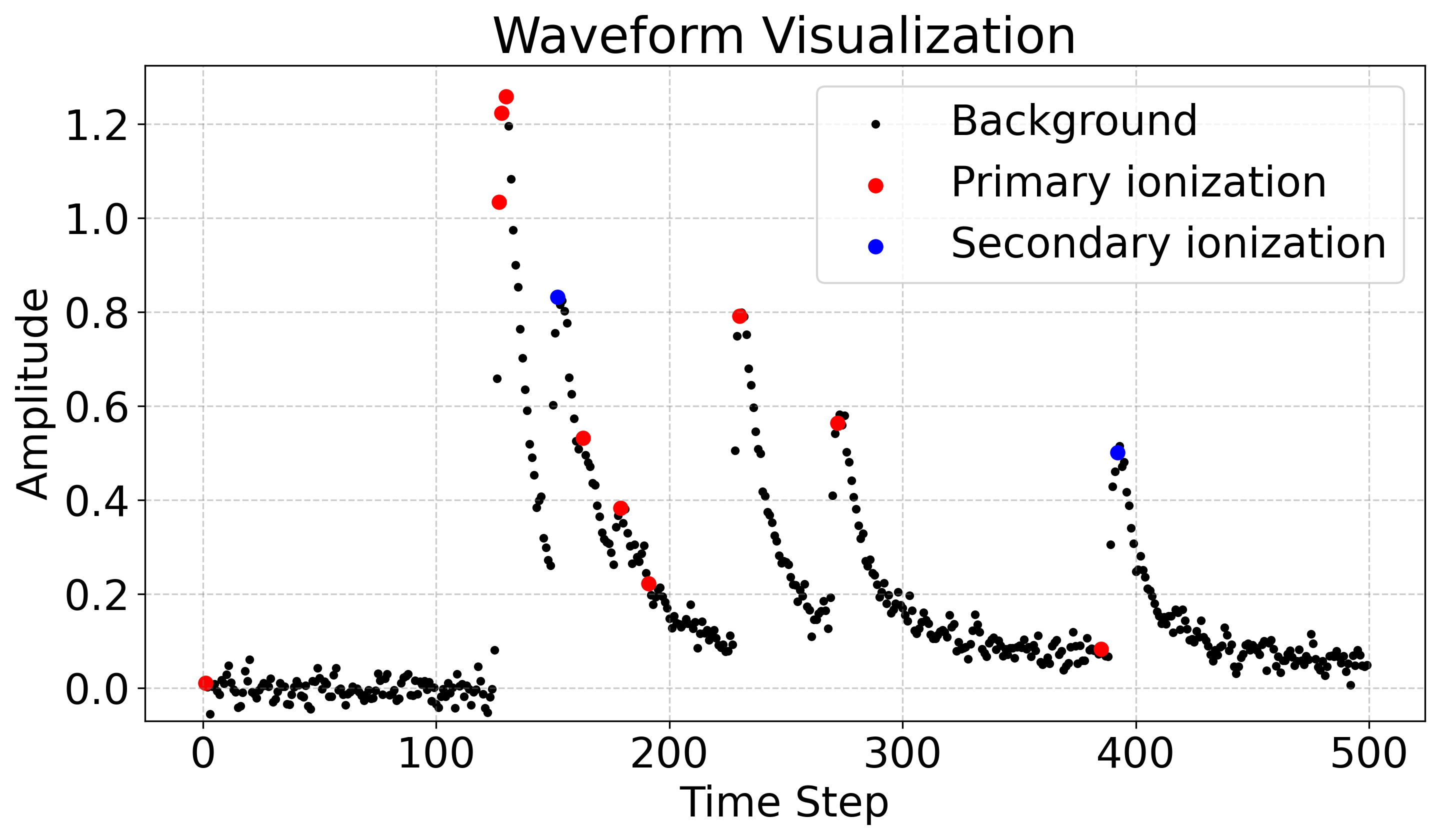}
    \caption{Example waveform from Ref.~\cite{Tian_2025}, including primary and secondary electron truth labels, truncated to 500 samples. 
    \label{fig:waveform}}
\end{figure}

Each waveform in this study is truncated to the first 500 samples (corresponding to a maximum drift time of approximately 350~ns) to emulate the reduction of the CEPC cell size expected for future drift chamber designs such as the IDEA detector~\cite{Dam:2025zed} at the Future Circular Collider (FCC)~\cite{Benedikt:2651299}.
Figure~\ref{fig:waveform} provides an example waveform used in the model development, with background, primary ionization, and secondary ionization truth labels provided. 

\subsection{Traditional cluster counting}


Traditional cluster counting approaches follow a two-step process: first applying peak finding algorithms to identify all candidate electron peaks, then merging adjacent peaks into single primary ionization clusters based on gas mixture properties and electron diffusion characteristics.

For peak-finding, the second-derivative-based (D2) algorithm is used in this work, which applies a second-order derivative to the raw waveform to identify signal peaks while maintaining a balance between detection efficiency and noise sensitivity. 
A baseline threshold $T_1$ (approximately three times the RMS noise amplitude) excludes background noise, while a second threshold $T_2$ is applied to the integrated second derivative for peak detection. Only positive derivatives corresponding to the rising edge are retained, as steep rising edges are characteristic features of ionization signals.

Following peak identification, two clustering algorithms are considered to merge electron peaks into primary ionization clusters. 
The Fixed-Clusterization Algorithm (FCA) uses fixed parameters across the entire waveform, including a time window size for forward-looking peak search, a normalization factor based on gas mixture cluster size characteristics, and a maximum number of peaks per window to limit noise and secondary ionization effects. 
The Adapted-Clusterization Algorithm (ACA) extends FCA by pre-dividing the waveform into time segments based on drift electron collection statistics, with each segment employing optimized parameter sets to adapt to time-dependent ionization physics such as cluster size variations with diffusion distance.

\subsection{ML-based direct cluster counting}


State-of-the-art ML models outperform classical methods for PID with cluster counting but often use computationally costly complex architectures, which are impractical under strict power and latency constraints. 
This work therefore focuses on fully connected (dense) neural networks, which are sufficiently small that no specific computing resources (eg. GPUs) are needed for model development.
The ML-based approach removes the need to do peak finding and cluster counting as separate steps, and instead does a direct regression of the number of clusters (eg. number of primary ionizations) for each track in a supervised way using the true number of clusters from Garfield++ as a label.  
The baseline model is a deep neural network (DNN) consisting of four layers following the 500 dimensional input: 8 units in the first, 32 in the second, 8 in the third, and a single output for cluster count regression. All dense layers use ReLU activation, with the output layer providing direct regression values.

The baseline model was trained for 75 epochs using the Adam optimizer with a learning rate of 0.0006. 400,000 events were used for training, 100,000 events were used for validation, and 100,000 events were used for testing. 
To calculate pion-kaon separation, the model trained over exclusively pion events was used to predict the number of clusters for both kaon and pion events, using 50,000 events for each momentum value. 
For each momentum point, we obtained distribution histograms for both particle types, extracted their respective means and standard deviations, and calculated the separation power $S$ scaled to a two meter track length.

To study real-time implementation, two compressed model variants were developed: pruned, which gradually removed 60\% of low-weight nodes over 25 epochs with minimal performance loss, and additionally a model with quantization-aware training via QKeras~\cite{qkeras}, which retrained the pruned model using 10-bit fixed-point precision (with 5 fractional bits).

\section{Results}

\begin{figure}[!htbp]
\centering
   \includegraphics[width=1.0\textwidth]{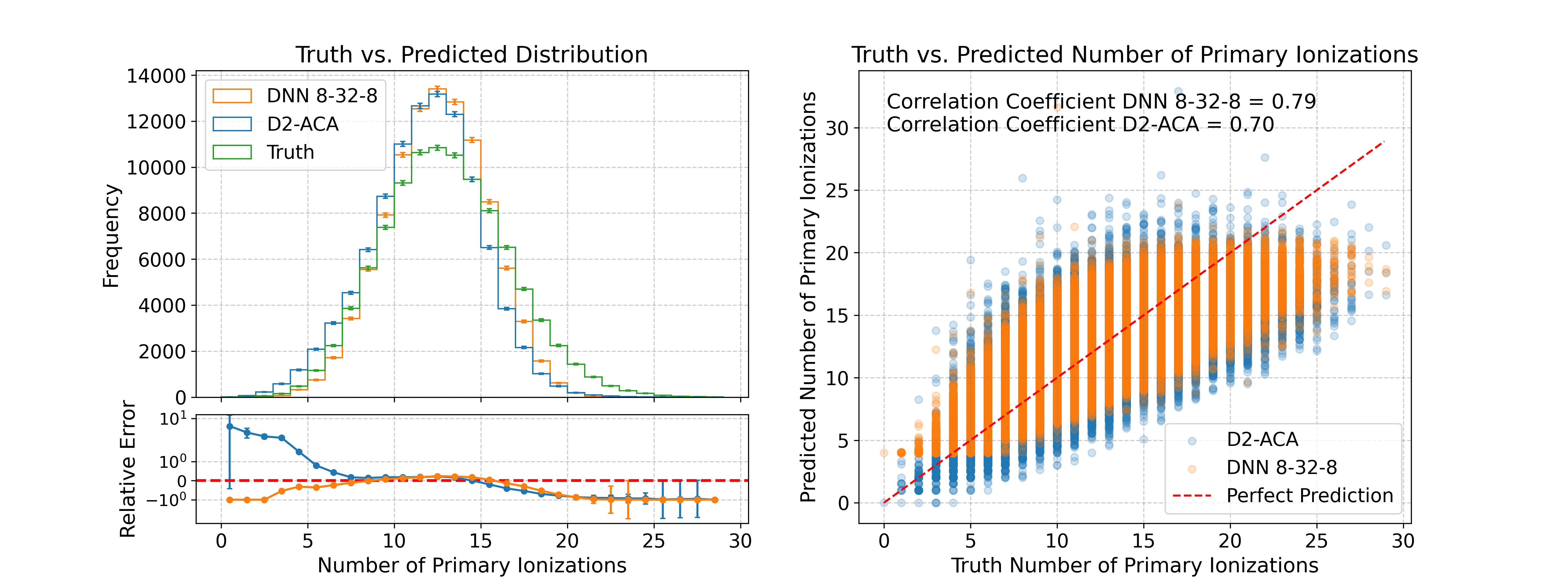}
    \caption{Performance of the baseline ML model for the task of predicting the truth number of clusters, comparing the predicted vs. truth cluster number distribution across the test set (left) and the correlation between the predicted and truth cluster number per test set event (right). 
    \label{fig:regressDistro_corr}}
\end{figure}

Figure~\ref{fig:regressDistro_corr} shows the capability of the baseline ML model to determine the predicted number of clusters for the test set, along with the correlation between predicted and truth cluster counts for each event, indicating good capability of the model to learn the cluster count for a given track. 

\vspace{-10pt} 



\textsc{} 

\begin{figure}[!htbp]
\centering
   \includegraphics[width=1\textwidth]{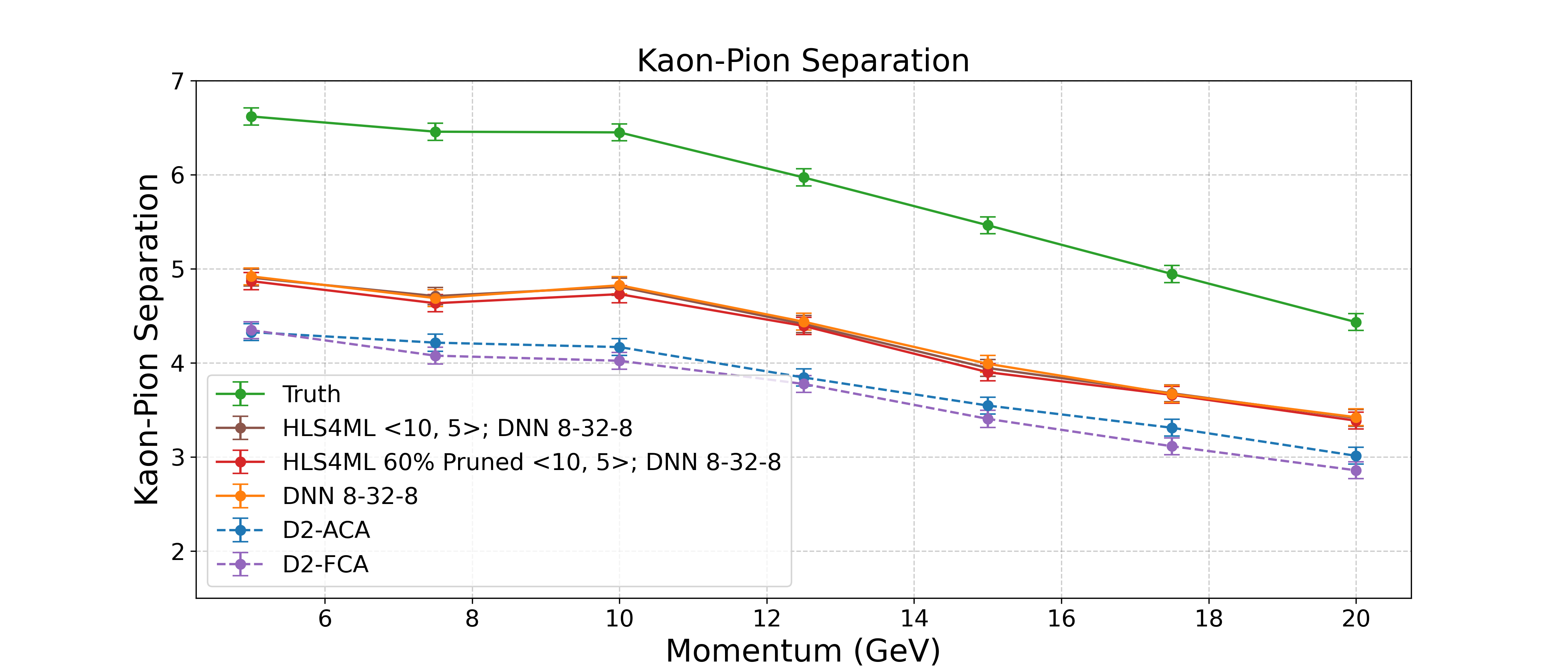}
    \caption{Performance of cluster counting methods for the task of separating pions and kaons. The ML-based methods outperform both traditional derivative-based methods across the full track momentum range, and the application of compression techniques can achieve a model that has $\mathcal{O}$(10) ns latency.  
    \label{fig:perf_pionKaon}}
\end{figure}

PID performance is evaluated using the pion-kaon separation power defined as: $S = \frac{|\frac{dN}{dx}_\pi - \frac{dN}{dx}_K|}{\frac{\sigma_\pi + \sigma_K}{2}}$ (taken from Ref.~\cite{Tian_2025}), where $\frac{dN}{dx}_{\pi(K)}$ and $\sigma_{\pi(K)}$ are the mean and standard deviation of primary ionizations per unit length for pions (kaons). 
The truncated dataset has an average of 12.5 clusters per waveform (in good agreement with the expected cluster density for the gas~\cite{Assran:2011ug} and the Garfield++ simulation), corresponding to a mean track length per cell of approximately 9.3~mm. 
To project the separation power for realistic track lengths, results are scaled from a single cell assuming a 2~m average particle trajectory. 
Figure~\ref{fig:perf_pionKaon} shows the obtained pion-kaon separation power over the test sets with track momentum range of [5, 20] GeV. 
A separation of $>$ 3$\sigma$ is achieved across the full tested track momentum range for an assumed 2~m track length. 

Trained models were synthesized using \texttt{hls4ml}~\cite{fastml_hls4ml, Duarte:2018ite} for FPGA implementation. 
This approach imagines a future target of embedded FPGAs for reconfigurable digital logic in a chip, which have been previously studied for front-end applications in high energy physics detectors~\cite{Gonski_2024}. 
To optimize for the latency constraint, the synthesis was configured with a reuse factor of 1 and \texttt{io\_parallel} option. 
The FPGA resource utilization for each algorithm is expressed in terms of Look-Up Tables (LUTs), Flip-Flops (FFs), and Digital Signal Processors (DSPs).

Table~\ref{tab:results} shows the achieved latency and FPGA resources for the baseline, pruned, and pruned+quantized models. 
The baseline direct regressor model achieves a total latency of 55 ns, putting these in the realm of real-time/trigger applications for future colliders with $\mathcal{O}$(10) ns bunch crossing rates. 
Pruning and quantization significantly reduce resource usage and thus on-chip power consumption, improving feasibility for front-end implementation.

Future studies into chip design and power studies are required to understand any further optimizations needed for realizable design, as well as codesign of analog and ML readout components. 
While this work presents the use of ML to perform front-end data compression by extracting the cluster count from the ionization waveform, a variety of other strategies exist that may prove superior given drift chamber readout specifications.
For example, data reduction could also be achieved by reducing the waveform to transmit the times of all primary ionizations, determining and transmitting PID directly from the front-end, or reducing the waveform data itself through autoencoder-based compression~\cite{Yue_2025}.
Additional work is needed to study all such options, which offer varying levels of compression, power, and data integrity, to determine which is best suited to future physics performance.

\begin{table}[h]
\centering
\begin{tabular}{ccccc}
\toprule
\textbf{Model} & \textbf{Latency [ns]} & \textbf{LUTs} & \textbf{FFs} & \textbf{DSPs}\\ 
\midrule
DNN 8-32-8 & 55 & 183,726 & 44,946 & 3,399 \\ 
HLS4ML Quantized <10,5> DNN 8-32-8 & 55 & 83,417 & 22,029 & 39 \\ 
HLS4ML 60\% Pruned; Quantized to <10,5> DNN & 45 & 127,818 & 26,002 & 19 \\
\bottomrule
\end{tabular}
\caption{FPGA resources (expressed in LUTs, DSPs, and FFs) and latency for the 3 ML regressor models considered.}
\label{tab:results}
\end{table}

\section{Conclusions}

A demonstration of ML-based cluster counting for drift chambers is presented. This work demonstrates the first ML approaches that have both improved pion-kaon separation with respect to traditional methods, as well as a clear path toward edge implementation given FPGA synthesis studies. 
Future work will focus on evaluating the power consumption of such designs and ensuring compatibility with the specifications of future drift chambers, and look into different on-chip implementations that afford a wider variety of data compression options. 
Code to reproduce the studies shown here can be found at \url{https://anonymous.4open.science/r/drift-chamber-ml-2DE0}.

\begin{ack}

The authors thank Charlie Young for helpful discussions. 
This work is supported by the U.S. Department of Energy under contract number DE-AC02-76SF00515 and the Office of the Vice Provost for Undergraduate Education at Stanford University.
\end{ack}

\bibliographystyle{unsrt} 
\bibliography{driftChamberML.bib}






\end{document}